\documentclass[onecolumn,12pt]{IEEETran}
\usepackage{bbm}
\usepackage{amsfonts}
\usepackage{tipa}
\usepackage{amssymb}
\usepackage{mathrsfs}
\usepackage[lined,boxed,commentsnumbered]{algorithm2e}
\usepackage{graphicx}
\usepackage{amsfonts,amssymb,amsmath}
\usepackage{latexsym}
\usepackage{multirow}
\usepackage{hyperref}
\usepackage{epsfig,epstopdf}
\newtheorem{lemma}{Lemma}
\newtheorem{proposition}{Proposition}

\begin{document}
\title{Digital Network Coding Aided Two-Way Relaying: Energy Minimization and Queue Analysis}
\author{Zhi~Chen~\IEEEmembership{Member,~IEEE,} Teng~Joon~ Lim~\IEEEmembership{Senior Member,~IEEE,}
        and~Mehul~Motani~\IEEEmembership{Member,~IEEE}
\thanks{Parts of this work were presented at IEEE ICCC (International Conf. on Communications in China) 2012, and submitted to IEEE ICCS (International Conf. on Communication Systems) 2012.
The authors are with the Department of Electrical and Computer Engineering, National University of Singapore, Singapore, 117583. Tel: +65 6601 2055.
Fax: +65 6779 1103.
Emails: \{elecz, eletj, motani\}@nus.edu.sg.}}

\maketitle

\maketitle
\baselineskip 24pt
\begin{abstract}\\
\baselineskip=18pt
In this paper, we consider a three node, two-way relay system with digital network coding.
The aim is to minimize total energy consumption while ensuring queue stability at all nodes,
for a given pair of random packet arrival rates.  Specifically, we allow for a set of transmission
modes and solve for the optimal fraction of resources allocated to each mode. First we formulate
and solve the static-channel problem, where all link gains are constant over the duration of
transmission. Then we solve the fading-channel problem, where link gains are random.
We call the latter the ergodic energy efficiency problem, and show that its solution has a water-filling
structure. Finally, we provide a detailed analysis of the queues at each node using a random scheduling
method that closely approximates the theoretical design, through a two-dimensional Markov chain model.
%
\end{abstract}

\IEEEpeerreviewmaketitle

\section{Introduction}
Network coding \cite{ahlswede2000network} has emerged as a viable means to improve throughput in complex networks. Messages at the packet level are linearly combined at intermediate nodes and forwarded to multiple intended destinations. Using knowledge of the manner in which messages were combined, communicated through some additional overhead bits, as well as knowledge of the message it contributed, a destination can successfully reconstruct the message intended for it. In this way, network throughput is greatly improved.

The two-way relay network \cite{liu2008network} exemplifies the use of network coding in digital communication. This simple network comprises two source nodes ($S_1$ and $S_2$) and one relay node ($R$), where $S_1$ and $S_2$ have information to exchange with each other. A direct link between $S_1$ and $S_2$ is unavailable. This model applies for instance to communication between a base station and a mobile user in cellular communications, in which the mobile user is in a location shadowed from the base station and coverage is provided in the area by means of a relay. Another typical application is that of satellite communication, where two satellites have messages for each other and can only communicate through a ground station.

With the aid of network coding, we can exchange two messages in two time slots with physical network coding \cite{Liew2006,zhang2009optimal,popovski2007physical} or three slots with digital network coding \cite{liu2008network,fong2011practical}, compared with the four slots needed with pure forwarding. With network coding, the relay receives messages from both source nodes, then combines these messages and broadcasts the network-coded packet to $S_1$ and $S_2$. The rate region for DNC-based two-way relay networks was first explored in \cite{liu2008network,fong2011practical} with various transmission modes including one-way forwarding and network-coded broadcasting. The former also investigated queue stability in the case of random data arrivals at each source node. The seminal work on physical layer network coding (PNC) presented in \cite{Liew2006} showed that two slots is sufficient for two packet transmissions provided that receiver-side SNR is sufficiently high and perfect synchronization could be achieved. \cite{Liew2011} and \cite{Zorzi2009asynchronous} discussed the practical design of asynchronous PNC systems. Multiuser detection was applied at the relay node to exploit asynchronous transmission in the uplink to form a network coded packet in the downlink.

In addition to throughput, resource allocation was investigated in \cite{Ruyet2011,jitvanichphaibool2009optimal,ho2008two,hausl2012resource}. In \cite{Ruyet2011}, optimal resource allocation with data fairness was discussed. Optimal resource allocation for analog network coding in a MIMO two-way relay network was investigated in \cite{zhang2009optimal}. In \cite{jitvanichphaibool2009optimal} and \cite{ho2008two}, optimal resource allocation for an ANC-based system was considered for OFDM systems. In \cite{hausl2012resource}, resource allocation was investigated under the scenario of asymmetric multi-way relay communication over orthogonal channels.

In \cite{li2010network}, the fading nature of a wireless channel was taken into consideration and the optimal position for the relay node was investigated. In \cite{lo2009network}, outage regions of DNC and ANC strategies were derived. It found that with the presence of a direct link, DNC was more promising than ANC in most cases in terms of outage performance.

In this paper, we formulate, simplify and solve the problem of allocation of channel resources to the various transmission modes in a DNC-based two-way relay network to minimize total energy usage in transmission. The constraint in the optimization problem is that queue stability is maintained at all four queues in the network for a given pair of average packet arrival rates $\lambda_1$ and $\lambda_2$ at $S_1$ and $S_2$, respectively. The scheduler, e.g., \ the relay node, is assumed to have full knowledge of the instantaneous channel coefficients in the four flat-fading links.

There are two scenarios to consider in such an energy minimization problem: (i) \underline{static channels}, which remain unchanged over the entire duration of the transmission; and (ii) \underline{fading channels}, which change with time so that over the duration of the transmission, all channel states are visited. For static channels, queue stability requires that throughput from $S_1$ ($S_2$) is at least $\lambda_1$ ($\lambda_2$) for a given set of channel gains; for fading channels, queue stability is guaranteed as long as throughput from $S_1$ averaged over the fading distribution is at least $\lambda_1$. In the fading case, the resulting optimization problem is solved by a water-filling procedure over the state space of the fading gains. This is tied to the solution of the static channel problem, which we will show to be equivalent to a convex optimization problem and hence is easily solved.

Assuming ergodicity in the channel processes, the fading-channel solution minimizes the long-term average energy used to deliver the arrival rate pair $(\lambda_1,\lambda_2)$. The derivation of this result is a natural complement to the first static-channel energy optimization problem.

Subsequently, we also design and simulate a random scheduling protocol based on the two designed resource allocation policies. To maintain finite queue length at each node, we introduce a back-off parameter $\epsilon$, so that the designed arrival rate is $\lambda_i(1+\epsilon)$ for an actual arrival rate of $\lambda_i$. This leads to the total energy usage being smaller than the objective function used in our design optimization. We investigate the behavior of the queues at the source nodes and the relay, for the random scheduling algorithm. Finding the average delay and queue length at $S_1$ and $S_2$ is straightforward, however the analysis of the two queues at the relay (for data from $S_1$ and $S_2$ respectively) required the use of a two-dimensional Markov chain. We present and verify through simulations this analysis in the second part of this paper.

The rest of this paper is organized as follows. In Section II, we introduce the energy usage optimization problem over static channels and provide the solution to this problem. In Section III, we introduce the ergodic energy usage optimization problem and present the associated solution with the water-filling structure. In Section IV, we present an energy efficient random scheduling protocol and provide the queueing analysis for this protocol. Simulation results are presented in Section V and Section VI concludes this work.

\emph{Notation}: Scalars are denoted by lower-case letters and bold-face lower-case letters are used for vectors and bold-face upper-case letters for matrices. In addition, we use \textbf{I} and $\bf{\Theta}$ to denote the identity matrix and the all-unity matrix respectively. $\textbf{1}$ is the row vector with all elements equal to unity.

\section{Static Channel Problem}
\begin{figure}[!t]
   \centering
   \includegraphics[width = 16cm]{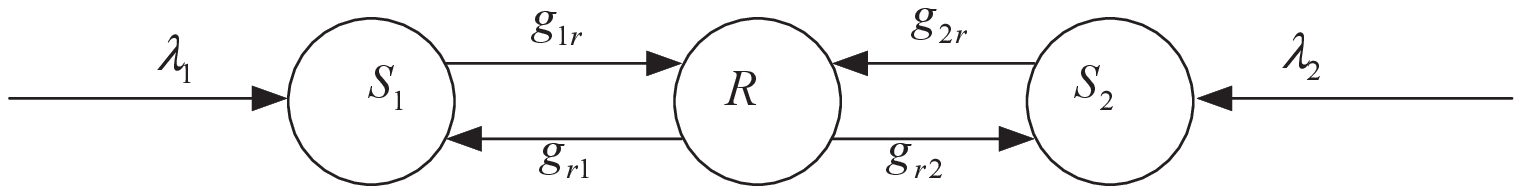}
   \caption{System model for a two-way relay network with random packet arrivals.} \label{fig:system}
   \end{figure}

In this section, we discuss the static channel problem, where the channel gains $g_{1r}$, etc.\ depicted in Figure \ref{fig:system} are deterministic, modelling the block-fading case, or simply the case of fixed channels as seen in wired systems.

\subsection{System Model and Problem Formulation}
Figure \ref{fig:system} depicts the two-way relay network of interest. Assume that packets arrive at $S_i$ according to a Poisson process at an average arrival rate of $\lambda_i$, $i \in \{1,2\}$, and we seek to maximize the energy efficiency of this system for a given average arrival rate pair $(\lambda_1,\lambda_2)$ while maintaining queue stability, by adjusting the fraction of time allocated to each of the following five transmission modes:
\begin{itemize}
\item {\em Mode 1:} $S_1$ transmits to $R$ at rate $R_1$ (uplink phase).
\item {\em Mode 2:} $S_2$ transmits to $R$ at rate $R_2$ (uplink phase).
\item {\em Mode 3:} $R$ broadcasts to $S_1$ and $S_2$ at broadcast rate $R_{3}$ (Network coding in downlink phase).
\item {\em Mode 4:} $R$ transmits only to $S_1$ at rate $R_{4}$ (one-way forwarding in downlink phase).
\item {\em Mode 5:} $R$ transmits only to $S_2$ at rate $R_{5}$ (one-way forwarding in downlink phase).
\end{itemize}

The four channel (power) gains $g_{1r}$, $g_{2r}$, $g_{r1}$ and $g_{r2}$, correspond to modes 1, 2, 4 and 5 respectively\footnote{Note that whether $g_{ri} = g_{ir}$ is inconsequential in this work.}. For convenience, we define $g_1 = g_{1r}$, $g_2 = g_{2r}$, $g_3 = \min(g_{r1},g_{r2})$, $g_4 = g_{r1}$ and $g_5 = g_{r2}$ so that $g_i$ is the channel gain in Mode $i$.
Receiver noise for each mode is modeled by i.i.d.\ Gaussian random variables with zero mean and unit variance. The power transmitted in Mode $i$ is denoted $P_i$. In other words, the signal to noise ratio (SNR) at the receiver in Mode $i$ is
\begin{equation}
   \mathsf{SNR}_i = P_i g_i.
\end{equation}
To achieve a rate of $R_i$ in Mode $i$ therefore requires that $P_i$ be exponentially related\footnote{An SNR gap term $\Gamma$ can be inserted into the expression but this only changes the effective channel gain to $g_{i}/\Gamma$ and does not affect the following discourse.} to $R_i$, via
\begin{equation} \label{eq:PiRi}
   P_i(R_i) = \frac{2^{R_i} - 1}{g_i}, \quad i = 1,\ldots, 5.
\end{equation}
It should be noted that in Mode 3, the rate of $R_3$ is the rate transmitted to {\em each} of $S_1$ and $S_2$, due to the use of network coding.




Suppose that a fraction $f_i$, $i = 1,\ldots,5$, of each time slot is allocated to Mode $i$ respectively. Therefore $f_i P_i(R_i)$ is proportional to the energy used for transmission in Mode $i$, if we disregard the possibility that queues may be empty. To minimize the total energy consumption per time slot while satisfying queue stability constraints therefore requires the solution of optimization problem {\bf P1}:
\begin{eqnarray}
\min_{f_i,R_i} && \sum_{i=1}^{5}f_i P_i(R_i)  \label{opt}
\end{eqnarray}
subject to the queue stability constraints
\begin{eqnarray}
\lambda_1 &\le& \min(f_1R_1,f_3R_3+f_5R_5)  \label{opt_1}\\
\lambda_2 &\le& \min(f_2R_2,f_3R_3+f_4R_4)  \label{opt_3}
\end{eqnarray}
and the physical constraints $\sum_{i=1}^5 f_i \leq 1$ and $f_i \geq 0$. In the next section, we show that {\bf P1} can be transformed into a simpler convex optimization problem with only $f_i$ as the design parameters and hence can be solved easily.

Finally, the queue lengths (in terms of packets) at $S_1$ and $S_2$ at epoch $t$ are denoted by $Q_1(t)$ and $Q_2(t)$. We also define $Q_{r1}(t)$ and $Q_{r2}(t)$ as the lengths of the two queues maintained at the relay node for $S_1$ and $S_2$, respectively.

\subsection{Problem Solution}
Problem {\bf P1} may be simplified through the following lemmas.
\begin{lemma} \label{lemma:3}
Under the optimal solution to {\bf P1}, we have
\begin{equation}
\left\{
\begin{array}{ll}
f_4 = 0, \hspace{0.2cm} \mbox{if} \hspace{0.5cm} \lambda_1 > \lambda_2\\
f_5 = 0, \hspace{0.2cm} \mbox{if} \hspace{0.5cm}\lambda_1 < \lambda_2
\end{array}
\right.\label{Eqn:lemma2}
\end{equation}
and $f_4 = f_5 = 0$ if $\lambda_1 = \lambda_2$.
\end{lemma}

\begin{IEEEproof}
Denote the optimal values of $f_i$ and $R_i$ by $f_i^*$ and $R_i^*$. Constraints (\ref{opt_1}) and (\ref{opt_3}) dictate that
\begin{eqnarray}
  f_3^* R_3^* + f_5^* R_5^* &\ge& \lambda_1 \\
  f_3^* R_3^* + f_4^* R_4^* &\ge& \lambda_2
\end{eqnarray}
Assume that $f_3^* R_3^* + f_4^* R_4^* - \lambda_2 = \epsilon > 0$. Then reducing $f_4^*$ by $\epsilon/R_4$ while keeping all other $f_i^*$ and all $R_i^*$ fixed results in constraint (\ref{opt_3}) still being satisfied, with a lower total energy. Therefore we can assume that $f_3^* R_3^* + f_4^* R_4^* = \lambda_2$. Similarly, $f_3^*R_3^* + f_5^*R_5^* = \lambda_1$.

Suppose that $f_4^* > 0$. If we reduce $f_4^*$ by $\delta_4$, constraint (\ref{opt_3}) requires that $f_3^*$ be increased by $\delta_4 R_4^*/R_3^*$. This in turn means that $f_5^*$ can be reduced by $\delta_4 R_4^*/R_5^*$. In other words, queue stability is maintained under the following adjustment:
\begin{eqnarray*}
  f_4^* &\rightarrow & f_4^* - \delta_4 \\
  f_3^* &\rightarrow & f_3^* + \delta_4 R_4^*/R_3^* \\
  f_5^* &\rightarrow & f_5^* - \delta_4 R_4^*/R_5^*
\end{eqnarray*}
The total energy used is now reduced by
\begin{eqnarray} \label{eq:DeltaE}
   \Delta E &=& \delta_4\left(\frac{R_4^*}{R_5^*}P_5^* + P_4^* - \frac{R_4^*}{R_3^*}P_3^*\right)
\end{eqnarray}
where $P_i^*$ is shorthand for $P_i(R_i^*)$. Note that $R_3^*$, $R_4^*$ and $R_5^*$ need only satisfy two equations equivalent to
\begin{eqnarray} \label{eq:f5R5}
   f_5^*R_5^* - f_4^*R_4^* &=& \lambda_1 - \lambda_2 \\ \nonumber
   f_3^*R_3^* + f_5^*R_5^* &=& \lambda_1.
\end{eqnarray}
We can thus always find a point in the solution space of (\ref{eq:f5R5}) to make the term in brackets in (\ref{eq:DeltaE}) positive. So as long as $\delta_4 > 0$ does not lead to a violation of (\ref{eq:f5R5}), we can reduce total energy by $\Delta E > 0$. If $\lambda_1 > \lambda_2$, this result and (\ref{eq:f5R5}) show that we should have $f_4^* = 0$, and $f_5^*R_5^* = \lambda_1 - \lambda_2$.

Similarly, if $\lambda_1 < \lambda_2$, then $f_5^* = 0$ and $f_4^*R_4^* = \lambda_2 - \lambda_1$; and if $\lambda_1 = \lambda_2$, then $f_4^* = f_5^* = 0$.
\end{IEEEproof}

A corollary of Lemma \ref{lemma:3} is that, since either one or both of $f_4^*$ and $f_5^*$ must be 0, $f_3^*$ cannot be zero, i.e.\ the broadcast phase (Mode 3) must be used. This is intuitive because one bit transmitted in Mode 3 is equivalent to two bits delivered, and hence we should devote as much resources as possible to Mode 3. The following lemma links $R_i$ to $f_i$.

\begin{lemma}\label{lemma:4}
The solution to \textbf{P1} satisfies
\begin{eqnarray}
R_i^* &=& \frac{\lambda_i}{f_i^*},\quad i = 1,2\label{3_1}\\
R_3^* &=& \frac{\min(\lambda_1,\lambda_2)}{f_3^*}\label{3_3}\\
R_4^* &=& \max\left(\frac{\lambda_2-\lambda_1}{f_4^*},0\right) \label{3_4}\\
R_5^* &=& \max\left(\frac{\lambda_1-\lambda_2}{f_5^*},0\right) \label{3_5}
\end{eqnarray}
\end{lemma}

\begin{IEEEproof}
By reducing $f_1$ and/or $R_1$, total energy is reduced, and so the energy-minimization problem must be solved when $f_1R_1$ is at its smallest value, i.e. $f_1^*R_1^* = \lambda_1$. Similarly, $f_2^*R_2^* = \lambda_2$.

Equations (\ref{3_4}) and (\ref{3_5}) arise directly from the proof of Lemma \ref{lemma:3}. Equation (\ref{3_3}) comes from noting that if $\lambda_1 < \lambda_2$, Lemma \ref{lemma:3} states that $f_5^* = 0$ and hence $f_3^*R_3^* = \lambda_1$ while if $\lambda_1 > \lambda_2$, $f_3^*R_3^* = \lambda_2$.
\end{IEEEproof}

Lemma \ref{lemma:4} implies that $\{R_i\}$ can be removed as optimization variables, leaving only $\{f_i\}$ in an equivalent optimization problem. Lemma \ref{lemma:3} (and also Lemma \ref{lemma:4}) implies that when $\lambda_1 > \lambda_2$, Mode 4 (forwarding from the relay to $S_1$) is not necessary, and that when $\lambda_1 < \lambda_2$, Mode 5 is not necessary. Hence there is no loss of generality in assuming that $\lambda_1 < \lambda_2$ from this point on, so that Mode 5 is no longer discussed and the min and max functions need not be invoked in Lemma \ref{lemma:4}. We now have
\begin{equation}
   R_3^* = \frac{\lambda_1}{f_3^*},\;\; R_4^* = \frac{\lambda_2 - \lambda_1}{f_4^*} \mbox{ and } R_5^* = 0.
\end{equation}

By substituting the new expressions for $\{R_i\}$ into {\bf P1}, assuming $\lambda_1 < \lambda_2$, we get the equivalent optimization problem {\bf P2}:
\begin{eqnarray}
    \min_{f_i} && f_1P_1\left(\frac{\lambda_1}{f_1}\right)+f_2P_2\left(\frac{\lambda_2}{f_2}\right)+f_3P_3\left(\frac{\lambda_1}{f_3}\right)\nonumber\\
    &&+f_4P_4\left(\frac{\lambda_2-\lambda_1}{f_4}\right) \label{opt2}\\
     \mbox{s.t.} && f_i \geq 0 \\
         && \sum_{i=1}^4 f_i \leq 1  \label{Eqn:unity}
\end{eqnarray}
where the power functions $P_i(\cdot)$ are given by (\ref{eq:PiRi}).

It is not difficult to show that the cost function in {\bf P2} is convex in $\{f_i\}$, and clearly the constraint set is also convex. Therefore the original problem {\bf P1} has been turned into an equivalent convex optimization problem {\bf P2}.

It should be noted too that the cost function in {\bf P2} is monotonically decreasing in $\{f_i\}$ and hence its solution must lie on the boundary of the constraint set. Since $f_i = 0$ for all $i$ is clearly not a viable solution, it must be that $\sum_i f_i = 1$ i.e., that all available time resources are fully utilized. This is easily understandable since increasing any $f_i$ leads to a corresponding reduction in the associated $R_i$, which means an exponential decrease in required power and hence we would want $f_i$ to be as large as possible.

Since {\bf P2} is a constrained convex optimization problem, its solution is found by solving the the Karush-Kuhn-Tucker (KKT) equations, which are easily derived as
\begin{eqnarray}
\frac{2^{\frac{\lambda_1}{f_1^*}}(1-\frac{\lambda_1 \ln 2}{f_1^*})}{g_{1}}+\beta^*-\frac{1}{g_{1}} &=& 0  \label{Eqn:1}\\
\frac{2^{\frac{\lambda_2}{f_2^*}}(1-\frac{\lambda_2 \ln 2}{f_2^*})}{g_{2}}+\beta^*-\frac{1}{g_{2}} &=& 0  \label{Eqn:2}\\
\frac{2^{\frac{\lambda_1}{f_{3}^*}}(1-\frac{\lambda_1 \ln 2}{f_3^*})}{g_{3}}+\beta^*-\frac{1}{g_{3}} &=& 0 \label{Eqn:3} \\
\frac{2^{\frac{\lambda_2-\lambda_1}{f_4}^*}(1-\frac{(\lambda_2-\lambda_1) \ln 2}{f_4^*})}{g_{4}}+\beta^*-\frac{1}{g_{4}} &=& 0 \label{Eqn:4} \\
\sum_{i=1}^4 f_i^* - 1 &=& 0 \label{Eqn:5}
\end{eqnarray}
where $\beta^*$ is the Lagrange multiplier.
Since (\ref{Eqn:1})--(\ref{Eqn:4}) are transcendental equations, it is infeasible to obtain explicit solutions in general and numerical methods are instead employed to obtain the solution to problem \textbf{P2}.

\subsection{Discussion and Insights}
Firstly, we give a lemma that links each active mode (i.e., those for which $f_i^*>0$) to the optimal solution to {\bf P2}.
\begin{lemma}\label{beta}
Under the optimal solution to {\bf P2}, we have for each active mode that
\begin{equation}
\frac{2^{R_i^*}(1-R_i^* \ln 2)}{g_{i}}+\beta^*-\frac{1}{g_{i}} = 0.  \label{eq: beta1}
\end{equation}
\end{lemma}
Note that this result immediately follows from equations (\ref{Eqn:1})-(\ref{Eqn:4}) and therefore the proof is omitted.

Based on Lemma \ref{beta},  some observations can be made and are presented in the following proposition.
\begin{proposition}\label{Lemma:5}
The optimal transmit rates and optimal transmit powers in respect of channel gains for active modes (i.e., $f_i^*,f_j^*>0$) are related as follows:
\begin{eqnarray}
&&R_i^*>R_j^* \hspace{0.2cm} \mbox{   and   } \hspace{0.2cm} P_i^*<P_j^*  \quad  \Leftrightarrow \quad g_i>g_j \label{eq:unequal}\\
&&R_i^*=R_j^* \hspace{0.2cm} \mbox{   and   } \hspace{0.2cm} P_i^*=P_j^*  \quad \Leftrightarrow \quad g_i=g_j \label{eq:equal}
\end{eqnarray}
where $i,j= 1,2,3,4,5$. In other words, $R_i^* - R_j^*$ has the same sign as $g_i - g_j$ whereas $P_i^* - P_j^*$ has the opposite sign.
\end{proposition}
\begin{IEEEproof}
Let us multiply by $g_i$ in (\ref{eq: beta1}), we then have the following equality for each active mode,
\begin{equation}
2^{R_i^*}(1-R_i^* \ln 2)+\beta^*g_i-1 = 0. \label{equality1}
\end{equation}

 Note that $R_i^*$ is an implicit function of $g_i$. Taking the derivatives on both sides in (\ref{equality1}) with respect to $g_i$ we obtain
\begin{equation}
 - 2^{R_i^*} (R_i^* \ln2)^2\frac{dR_i^*}{dg_i} +\beta^*=0,  \label{equality2}
\end{equation}
from which it follows that
\begin{equation}
\frac{dR_i^*}{dg_i} = \frac{\beta^*}{2^{R_i^*} (R_i^* \ln2)^2}>0.\label{equality3}
\end{equation}
The inequality comes from the fact that
$R_i^*$ for active modes and $\beta^*$ are positive values. Hence $R_i$ is a monotonically increasing function of $g_i$.

Recalling that $R_i^* = \log_2(1+P_i^*g_i)$, (\ref{equality1}) can also be written as
\begin{equation}
(P_i^*g_i+1)(1-\ln(1+P_i^*g_i))+\beta^*g_i-1 = 0. \label{eq: beta2}
\end{equation}

Taking derivatives with respect to $g_i$ on both sides of (\ref{eq: beta2}) we arrive at
\begin{equation}
-\ln(1+P_i^*g_i)(P_i^*+g_i\frac{dP_i^*}{dg_i}) +\beta^*=0. \label{equality4}
\end{equation}

With some manipulations we can obtain (\ref{equality5})-(\ref{equality7}) below.
\begin{eqnarray}
 \frac{dP_i^*}{dg_i}&=& \frac{\beta^*-P_i^*\ln(1+P_i^*g_i)}{g_i\ln(1+P_i^*g_i)}\label{equality5}  \\
 &=&\frac{\ln(1+P_i^*g_i)-P_i^*g_i}{g_i^2\ln(1+P_i^*g_i)}\label{equality6}  \\
 &<& 0  \label{equality7}
\end{eqnarray}
where (\ref{equality6}) comes directly from (\ref{eq: beta2}) and (\ref{equality7}) comes from the fact that $\ln(1+x)<x$ if $x>0$.
Hence $P_i^*$ is a strict decreasing function of $g_i$.

Considering both (\ref{equality3}) and (\ref{equality7}), Proposition \ref{Lemma:5} is proved.
\end{IEEEproof}

It is interesting that Proposition \ref{Lemma:5} implies that the relative values of $R_i^*$ and $P_i^*$ depend only on the relative values of the corresponding channel gains, but not at all on the average arrival rates.
The optimal time fraction for each mode, however, does relate to the corresponding arrival rate in the form $f_i^*=\frac{\lambda_i}{R_i^*}$.

Note also that Proposition \ref{Lemma:5}, which relates the optimal power, rate and link gain in each mode, is useful for initializing the power and rate optimization routine so that the iterative procedure converges more quickly. For instance, if $g_1 > g_2$, we should initialize $P_2$ to exceed $P_1$, and $R_1$ to exceed $R_2$.

In addition, for the case that $\lambda_1$ and $\lambda_2$ are very small ($\lambda_i \ll 1$), we can use a Taylor series to approximate the exponential functions and obtain the closed form power allocation solution for each mode, with $\beta$ obtained through a one-dimensional bisection search:
\begin{eqnarray}
  f_1^* &=& \lambda_1 \left(1+\beta^*-\frac{1}{g_{1}}\right)^{\frac{1}{2}} \label{Eqn:appro1}\\
  f_2^* &=& \lambda_2 \left(1+\beta^*-\frac{1}{g_{2}}\right)^{\frac{1}{2}}  \label{Eqn:appro2}\\
  f_3^* &=& \lambda_1 \left(1+\beta^*-\frac{1}{g_{3}}\right)^{\frac{1}{2}}  \label{Eqn:appro3}\\
  f_4^* &=& (\lambda_2 - \lambda_1)\left(1 + \beta^* - \frac{1}{g_{4}}\right)^{\frac{1}{2}}
 \\
  f_4^* &=& 1 - f_1^* - f_2^* - f_3^*.
\end{eqnarray}

We should also emphasize that when $\lambda_1 > \lambda_2$, the last two terms in the cost function of {\bf P2} change to
$$
   f_3P_3\left(\frac{\lambda_2}{f_3}\right) + f_5P_5\left(\frac{\lambda_1 - \lambda_2}{f_5}\right)
$$
and the problem can be solved with the appropriate substitutions. Hence there is no loss in generality in the assumption that $\lambda_1 < \lambda_2$ that we used for the majority of this section.


\section{Maximizing Ergodic Energy Efficiency in Fading Channels}
Unlike in the previous section, assume now that $g_i$, $i = 1,\ldots,5$, are random variables with known density functions $p(g_i)$. Assuming ergodicity in the random processes $g_i(t)$, where $t$ represents time, minimization of average total transmit energy in the long run is formulated as {\bf P3}:
\begin{eqnarray}
   \min_{f_i,R_i(g_i)} && \sum_{i=1}^{5}f_i \bar{P}_i  \label{opt}
\end{eqnarray}
subject to
\begin{eqnarray}
\lambda_1 &\le& \min(f_1\bar{R}_1,f_3\bar{R}_3 + f_5\bar{R}_5)  \label{lopt_1}\\
\lambda_2 &\le& \min(f_2\bar{R}_2,f_3\bar{R}_3 + f_4\bar{R}_4)  \label{lopt_2}\\
\sum_{i=1}^5 f_i &\leq& 1.
\end{eqnarray}
In the above
$$
   \bar{P}_i = E[P_i(R_i(g_i))] = \int_0^\infty P_i(R_i(g_i))p(g_i) dg_i,
$$
and
$$
   \bar{R}_i = E[R_i(g_i)] = \int_0^\infty R_i(g_i)p(g_i) dg_i,
$$
where $\bar{P}_i$ and $\bar{R}_i$ are averaged over the channel gain distributions. We term problem {\bf P3} an {\em ergodic energy efficiency maximization} problem. Note that the minimization is over $(f_i,R_i(g_i))$ once the functional form of $R_i(g_i)$ is found, we obtain $P_i(R_i(g_i))$ from (\ref{eq:PiRi}) and hence $\bar{P}_i$, while $\bar{R}_i$ is obtained through averaging $R_i(g_i)$ over $g_i$.

In other words, solving problem {\bf P3} yields a set of time fractions $\{f_i^*\}$ that minimizes an upper bound on the long-term average energy used, while guaranteeing long-term queue stability, assuming that the scheduler has knowledge of the instantaneous channel gains $g_i$, $i = 1,\ldots,5$, as well as the distribution of $g_i$. At the same time, we obtain a rate allocation $R_i^*(g_i)$ dependent on the instantaneous channel gains $g_i$.

It can be verified that Lemmas 1--2 still hold for this optimization problem and hence it can be translated into another equivalent optimization problem, as shown presently.

Again assuming without loss of generality that $\lambda_1<\lambda_2$, {\bf P3} reduces to \textbf{P4}:
\begin{eqnarray}
\min_{f_i,R_i(g_i)} && \sum_{i=1}^{4}f_i \bar{P}_i  \label{opt}
\end{eqnarray}
subject to
\begin{eqnarray}
\lambda_1 &=& f_1\bar{R}_1 = f_3\bar{R}_3  \label{lopt_1}\\
\lambda_2 &=& f_2\bar{R}_2 \label{lopt_2}\\
\lambda_2-\lambda_1 &=& f_4\bar{R}_4 \label{lopt_1}\\
\sum_{i=1}^4 f_i &\leq& 1.
\end{eqnarray}

The Lagrangian function corresponding to {\bf P4} is
\begin{equation}
   F = \sum_{i=1}^4 f_i \bar{P}_i - \sum_{i=1}^4\beta_i(f_i\bar{R}_i -\lambda_i)+\gamma \left(\sum_{i=1}^4 f_i - 1\right)
 \end{equation}
where we used the notation $\lambda_3=\lambda_1$ and $\lambda_4=\lambda_2-\lambda_1$ to denote virtual arrival rates for Mode 3 and Mode 4 respectively.

Since {\bf P4} is a convex optimization problem, its solution is given by the KKT conditions:
 \begin{eqnarray}
 \bar{P}_i^* - \beta^*_i\bar{R}^*_i + \gamma^* &=& 0, \quad 1\le i \le 4  \label{eqn:kkt1}\\
 f^*_i\bar{R}_i^* - \lambda_i &=& 0, \quad 1 \le i \le 4 \label{eq:kkt1a}\\
 f_i^*\frac{2^{R^*_i}}{g_i} \ln 2 - f^*_i\beta^*_i &=& 0,   \quad 1\le i \le 4   \label{eqn:kkt2}\\
 \beta^*_i, \gamma^* &\geq& 0,  \quad 1\le i \le 4 \label{eqn:kkt3}\\
  \sum_{i=1}^4 f^*_i &=& 1
 \end{eqnarray}
where the asterisks denote optimal values.

It can be found from (\ref{eqn:kkt2}) that $\frac{2^{R_i^*}}{g_i}=\beta^*_i \log_2 e$ and recall the power-rate equality in (\ref{eq:PiRi}) we have
\begin{equation}
 P_i^*(R_i^*(g_i)) = \left[\beta^*_i \log_2 e -\frac{1}{g_i}\right]^+  \label{eqn:beta}
\end{equation}
Then by definition,
\begin{eqnarray}
\bar{R}^*_i &=& \int_{\frac{1}{\beta^*_i \log_2 e}}^\infty \log_2(1+P_i^*(R^*_i)g_i)p(g_i) dg_i\\
&=& \int_{\frac{1}{\beta^*_i \log_2 e}}^\infty \log_2(\, \beta^*_ig_i \log_2 e  \,)p(g_i)dg_i \label{eqn:optR}
\end{eqnarray}

\begin{figure}[t]
   \centering
   \includegraphics[width = 7.8cm]{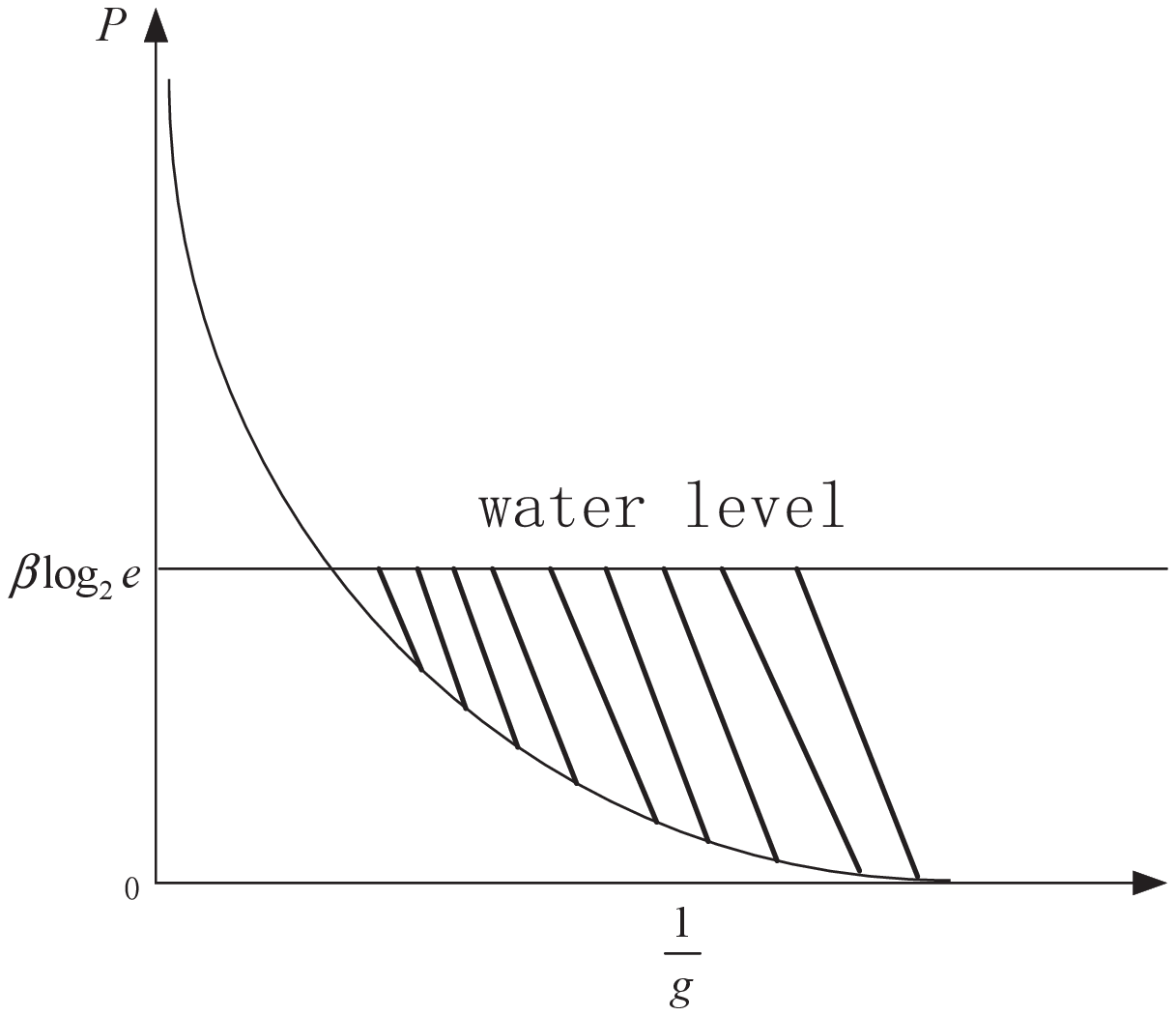}
   \caption{Water filling diagram for each link.} \label{fig:water_filling}
   \end{figure}

Therefore the power allocation solution over each link is of the form of water-filling and the illustrative diagram is shown in Fig. \ref{fig:water_filling}.

Note that $\bar{P}^*_i$ can also be found in terms of $\beta^*_i$ and $\gamma^*$. The optimal $f_i^*$ is given by (\ref{eq:kkt1a}). Finally, we obtain $\beta_i^*$ and $\gamma^*$ from (\ref{eqn:kkt1}) by performing a multidimensional bisection search method to make $\beta^*_i\bar{R}_i - \bar{P}_i$ equal (to $\gamma^*$) and positive for all $i$.

It is interesting to note that similar observations to Proposition \ref{Lemma:5} can be made for the ergodic energy optimization problem for Rayleigh fading channels. They are given in the following proposition. The detailed proof however is left in the appendix.
\begin{proposition}\label{Lemma:7}
Under the optimal solution to ergodic energy consumption optimization problem over Rayleigh fading channels, the relation of the optimal transmit power and rate of each mode with the associated average link gain can be given by,
\begin{eqnarray}
&&\bar{P}_i^*<\bar{P}_j^*   \quad\mbox{and}\quad \bar{R}_i^*>\bar{R}_j^*   \quad   \Leftrightarrow \quad \bar{g}_i>\bar{g}_j\\
&&\bar{P}_i^*=\bar{P}_j^*   \quad\mbox{and}\quad \bar{R}_i^*=\bar{R}_j^*    \quad \Leftrightarrow \quad
 \bar{g}_i=\bar{g}_j
\end{eqnarray}
where $i,j= 1,2,3,4,5$ if $f_i^*,f_j^*>0$. In other words, $\bar{P}_i^*$ is a decreasing function of $\bar{g}_i$ whereas $\bar{R}_i^*$ is an increasing function of $\bar{g}_i$ in the ergodic case as well.
\end{proposition}

\section{Scheduling Protocol: Queuing and Actual Energy Efficiency}
We now consider scheduling strategies which make use of the optimal time-sharing fractions $\{f_i\}$ (and hence constant rates $\{R_i\}$ for {\bf P1} and varying rate for {\bf P3}) for each transmission mode when solving energy minimization problems {\bf P1} and {\bf P3} respectively. 
We then describe below an intuitive protocol called the energy-efficient random scheduling protocol (EERSP).
%
%
\begin{enumerate}
   \item At the start of a packet interval of duration $T$, the relay randomly chooses a transmission mode with probability $P[\mbox{Mode i}] = f_i$.

   \item If Mode 1 is selected, the relay will inform $S_1$ to transmit.  If Mode 2 is selected, the relay will inform $S_2$ to transmit.  The selected source will transmit if its data buffer is non-empty.

   \item If Mode 3 is selected, the relay will transmit.  If both queues at $R$ are non-empty (i.e., $\min(Q_{r1}(\cdot),Q_{r2}(\cdot))>0$), it will broadcast network coded packets to both sources. If only $Q_{r1}(\cdot)>0$ (or only $Q_{r2}(\cdot)>0$), the relay will forward data to $S_1$ (or $S_2$). If both queues at $R$ are empty, it will remain silent.

   \item If Mode 4 is selected, the relay will forward packets to $S_1$ if $Q_{r1}(\cdot)>0$.  If Mode 5 is selected, the relay will forward packets to $S_2$ if $Q_{r2}(\cdot)>0$.
\end{enumerate}

\subsection{Queue Analysis for {\bf P1}}
We now analyze the queuing performance of EERSP. The data arrival processes at both $S_1$ and $S_2$ are assumed to be Poisson. For brevity, we shall only focus on the queue length $Q_1(t)$ and $Q_{r2}(t)$. The analysis for $Q_2(t)$ and $Q_{r1}(t)$ is very similar to that for $Q_1(t)$ and $Q_{r2}(t)$ and is omitted for brevity.

Here we use a two-dimensional Markov chain to model the backlog states of $(Q_1(t),Q_{r2}(t) )$ and define $q_{(m,k),(i,j)}$ as the one-step transition probability of the event that $Q_{r2}(t+1)=j$ and $Q_1(t+1)=i$ given that $Q_{r2}(t)=k$ and $Q_1(t)=m$. There are four classes of queue transitions arising from the five transmission modes:

Class I: $S_1$ is selected for transmission in Mode 1;

Class II: $R$ is selected for transmission in Mode 3;

Class III: Mode 2 or Mode 4 is selected;

Class IV: $R$ is selected for transmission in Mode 5.

Without loss of generality, assume that one time slot of duration $T$ corresponds to one channel use, and that capacity is measured in units of packets. Hence, for Mode 1 transmission, where a rate $R_1^*$ is supported, we assume that a maximum of $R_1^*$ packets can be transmitted from $S_1$ to $R$. In general, in Mode i, a maximum of $R_i^*$ packets can be transmitted. The queue analysis below follows naturally from this view.

For Class I up to $R_1^*$, packets may be received reliably at $R$ from $S_1$, the conditional one-step transition probability is
\begin{equation}
q_{(m,k),(i,j)}=\left\{
\begin{array}{ll}
a_i, & \quad  m=0,j=k\\
a_{i}, & \quad  0<m<R_1^*, j=k+m\\
a_{i-m+R_1}, & \quad m \geq R_1^*, j =k+R_1^*\\
\end{array}
\right.\label{Eqn:scenario_1}
\end{equation}
where
\begin{equation}
a_j=\frac{(\lambda_1 T)^j}{j!}\exp(-\lambda_1 T)
\end{equation}
is the probability that $j$ packets arrive at $S_1$ within the current slot.
For the the case where $Q_1(t)=0$, no packets are queued for transmission by $S_1$, therefore $Q_{r2}(t+1)=Q_{r2}(t)$. For $Q_1(t+1)$ to be equal to $i$, there must have been $i$ arrivals at $S_1$ in the duration T of the $(t+1)$-st slot.

For the case where $0<Q_1(t)<R_1^*$, all queued packets at $S_1$ will be transmitted during slot $t+1$ and hence $Q_{r2}(t+1)=Q_{r2}(t)+Q_1(t)$. Therefore, $Q_1(t+1)$ is determined by the number of packets arrived during slot $t+1$.

For the case that $Q_1(t)>R_1^*$, we have
$Q_{r2}(t+1)=Q_{r2}(t)+R_1^*$ since $S_1$ can transmit at most $R_1^*$ packets during one slot. $Q_1(t+1)$ is thus determined both by the number of residual buffed packets $Q_1(t)-R_1^*$ and the number of new packet arrivals, i.e, $Q_1(t+1)=Q_1(t)-R_1^*+i$ provided that there are $i$ arrivals during slot $t+1$.
It should be noted that there are no other possible transitions.

Under Class II, $R$ will transmit network coded data to $S_2$ in Mode 3. In this scenario, queue length increment at $S_1$ is determined by the number of arrived packets at $S_1$ within this slot. For queueing at $R$ for $S_2$, there will be two cases after transmission. If $Q_{r2}(t)$ is less than $R_3^*$, it will be emptied at slot $t+1$. Otherwise, there will be still $Q_{r2}(t)-R_3^*$ packets buffered at $R$ for $S_2$. The conditional transition probability thus can be given as by
\begin{equation}
q_{(m,k),(i,j)}=\left\{
\begin{array}{ll}
a_{i-m}, & \quad  j=0, k \leq R_3^*\\
a_{i-m}, & \quad  j=k-R_3^*, k>R_3^*\\
\end{array}
\right.\label{Eqn:scenario_2}
\end{equation}

Under Class III, $Q_{r2}$ remains the same in the next slot. Only the external arrival at $S_1$ will change the state of $Q_1(t)$. The conditional transition probability thus is
\begin{equation}
q_{(m,k),(i,j)}=a_{i-m}, \quad j=k \label{Eqn:scenario_3}
\end{equation}

The analysis for Class IV is similar to that for Class II and is omitted for brevity.

\begin{figure}[!t]
   \centering
   \includegraphics[width = 7.8cm]{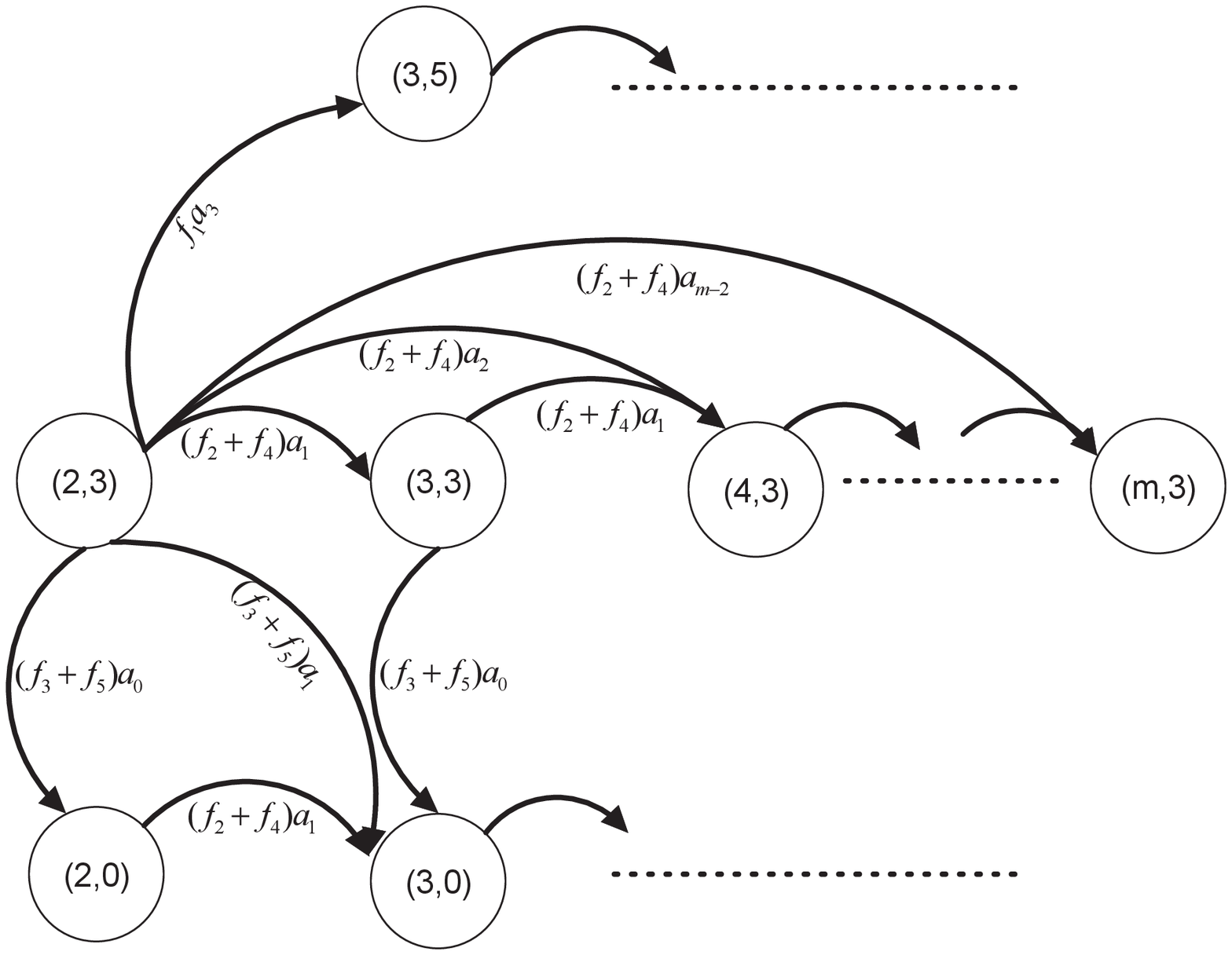}
   \caption{Two-dimensional Markov chain corresponding to EERSP protocol} \label{fig:markov}
   \end{figure}

Combining all these modes and current queue state, the state space and corresponding transition probability for $Q_1(t)$ and $Q_{r2}(t)$ of the EERSP protocol are
given below and is shown in Fig. \ref{fig:markov}. 

To simplify the analysis, the one-step transition probability matrix of the two-dimensional Markov chain can be written as that of a one-dimensional Markov chain, i.e., $\textbf{P}=\{ q_{mn+k,in+j} \}$ where $n$ is a large positive number and
$q_{(m,k),(i,j)}$ is represented by the element of $\textbf{P}_{mn+k,in+j}$. The stationary distribution thus is
\begin{equation}
\bf{\pi} = \textbf{1} \cdot (\textbf{I}-\textbf{P}+\bf{\Theta})^{-1}.  \label{Eqn:stationary}
\end{equation}

The average queue length at $S_1$ and that at $R$ for $S_2$ then are
\begin{eqnarray}
\overline{Q_1}&=&\sum_{i=0}^{\infty}\sum_{j=0}^{n}i\pi_{in+j} \label{queue1}\\
\overline{Q_{r2}}&=&\sum_{i=0}^{\infty}\sum_{j=0}^{n}j\pi_{in+j} \label{queue2}
\end{eqnarray}

\subsection{Queue and Delay Analysis for {\bf P3}}
The analysis of queueing for the fading channel problem is similar to the analysis of the static channel problem in the above subsection. The four classes of state transitions still apply, but the rates supported in each mode are no longer constant. They are instead random variables due to (\ref{eqn:beta}), which makes $R_i^*$ a function of $g_i$, a random variable. Therefore, we define $c_n$ as the probability that $S_1$ is only able to transmit $n$ packets in one slot if it is selected for transmission, while $r_n$ and $q_n$ are the probability that $R$ can transmit $n$ packets in one slot if it is selected for transmission in Mode 3 and Mode 5 respectively.

For Class I, packets may be received reliably at $R$ from $S_1$, the conditional one-step transition probability then is
\begin{equation}
q_{(m,k),(i,j)}=\left\{
\begin{array}{ll}
a_i, & \quad  m=0,j=k\\
a_{i} \cdot \sum_{n=m}^{\infty}c_n, & \quad  0<m, j=k+m\\
a_{i-m+n}c_n, & \quad 0 \leq n < m, j=k+n\\
\end{array}
\right.\label{Eqn:scenario_11}
\end{equation}
Where $a_i$ is as defined in the previous subsection.
The first term on the right hand side of (\ref{Eqn:scenario_11}) represents the case that $Q_1(t)$ is zero. As no packets are queued at $S_1$ for transmission, we have $Q_{r2}(t+1)=Q_{r2}(t)$ and $Q_1(t+1)$ is determined by the number of packets arrival in the duration T of the $t+1$st slot.  The second term accounts for the case that the instantaneous transmit rate $R_1^*(t)$ of $S_1$ is higher than $Q_1(t)$ and hence $Q_{r2}(t+1)=Q_{r2}(t)+Q_1(t)$. The probability that $R_1^*(t)>=Q_1(t)$ is given by $\sum_{i=Q_1(t)}^{\infty} c_i$. In this case, $Q_1(t+1)$ is also determined by the number of new arrivals during slot $t+1$ since packets buffered previously are totally delivered to $R$. For the third term where $R_1^*(t)<Q_1(t)$, we have $Q_2(t+1)=Q_2(t)+R_1^*(t)$ since only as most $R_1^*(t)$ packets can be transmitted during one slot. $Q_1(t+1)$ is hence given by the sum of the number of residual buffered packets and the number of new arrivals, i.e., $Q_1(t+1)=Q_1(t)-R_1^*(t)+i$ given there are $i$ arrivals during slot $t+1$.

For Class II, $R$ will transmit network coded data to $S_2$ in Mode 3. The conditional transition probability is given as follows.
\begin{equation}
q_{(m,k),(i,j)}=\left\{
\begin{array}{ll}
a_{i-m}\sum_{n=k}^{\infty}r_k, & \quad  j=0\\
a_{i-m}r_n, & \quad  j=k-n, 0 \leq n < k\\
\end{array}
\right.\label{Eqn:scenario_22}
\end{equation}
Where queue length increment at $S_1$ is determined by the number of arrived packets at $S_1$ within this slot. For queueing at $R$ for $S_2$, $Q_{r2}(t+1)$ is determined by current buffering length $Q_{r2}(t)$ and the instantaneous broadcasting rate of $R$, i.e., $R_3^*(t)$. If $Q_{r2}(t)<R_3^*(t)$, it will be emptied at slot $t+1$. Otherwise, there will be still $Q_{r2}(t)-R_3^*(t)$ packets buffered at $R$ for $S_2$.

Similarly, we can get the one-step transition probability for Class III and Class IV. Combining them together, the one-step transition probability of the EERSP protocol with ergodic energy optimization can be given as follows.
\begin{equation}
q_{(m,k),(i,j)}=\left\{
\begin{array}{ll}
a_i, & \quad m=k=j=0 \\
(1-f_1+f_1c_0)a_{i-m}, &\quad m>0,k=j=0\\
(f_3r_n + f_5q_n)a_{i-m}, & \quad  j=k-n,k>0,0 \leq n <k\\
a_{i-m}(f_3\sum_{n=m}^{\infty}r_n+f_5\sum_{n=m}^{\infty}q_n), & \quad  j=0, k>0\\
(f_2+f_4+f_1+f_3r_0+f_5q_0)a_i, & \quad j=k>0, m=0\\
(f_2+f_4+f_1c_0+f_3r_0+f_5q_0)a_{i-m}, & \quad j=k>0, m>0\\
f_1a_{i-m+n}c_n, & \quad m>0, j=k+n,0 \leq n \leq m-1\\
f_1a_{i} \cdot \sum_{n=m}^{\infty}c_n, & \quad  0<m, j=k+m\\
0, & \quad \mbox{else}\\
\end{array}
\right.\label{Eqn:queue_q1}
\end{equation}

In a similar manner, average queue length of $S_1$ and that of $R$ for $S_2$ can be characterized by using (\ref{Eqn:stationary})-(\ref{queue2}).

\subsection{Practical Energy Consumption with Positive $\epsilon$}
To ensure reasonable running times and memory requirements in simulation, we design for average arrival rates that are slightly larger than the actual rates, i.e., $\lambda_i \rightarrow (1+\epsilon)\lambda_i$ ($i=1,2$), where $\epsilon$ is a small positive value. In this case it should be noted that the actual consumed energy might be lower than the designed energy consumption with positive $\epsilon$ since there might be idle slots of each active mode in implementation. Therefore, it is interesting to investigate actual energy consumption in the regime of positive $\epsilon$ and this is what will be done in this section. Note that here we only focus on the constant link gain scenario. The practical energy consumption under the scenario of fading channel can be derived in a similar manner and is hence omitted.

Let $f_i^*$ be the optimal time fraction of each mode for \emph{virtual} arrival rate pair ($\lambda_1(1+\epsilon)$, $\lambda_2(1+\epsilon)$). Let $P_i^*$ be the optimal power level and $R_i^*$ the optimal transmit rate of Mode i for \emph{virtual} arrival rate pair. Let $\pi^1_{ij}$ be the stationary distribution for queue pair $(Q_1(t),Q_{r2}(t) )$ and $\pi^2_{ij}$ be the stationary distribution of queue pair $(Q_2(t),Q_{r1}(t) )$ respectively for \emph{actual} arrival rate pair. Considering the two exclusive scenarios that $Q_i(t)<R_i^*$ and that $Q_i(t)>R_i^*$ together, we can derive the actual energy consumption for each active mode respectively.

For Modes 1 and 2, the actual energy consumed can be given by,
\begin{equation}
\bar{E}_i^{'}=\sum_{k=R_i^*}^{\infty}\sum_{j=0}^{\infty}\pi^i_{kj}P_i^*+ \sum_{0}^{k=R_i^*-1}\sum_{j=0}^{\infty} \left( \pi^i_{kj} \cdot \frac{2^k-1}{g_i} \right)  \quad  i=1,2
\end{equation}
where the first term accounts for the case that current queue length is higher than transmit rate of this mode whereas the second term accounts for $Q_i(t) < R_i^*$.

Similarly, for Modes 4 or 5 whichever is active ($f_i^*>0$), the actual energy consumed is given by,
\begin{equation}
\bar{E}_i^{'}= \sum_{k=R_i^*}^{\infty}\sum_{j=0}^{\infty}\pi^{6-i}_{jk}P_i^*+ \sum_{0}^{k=R_i^*-1}\sum_{j=0}^{\infty} \left( \pi^{6-i}_{jk} \cdot \frac{2^k-1}{g_i} \right) \quad i=4(5)
\end{equation}

For Mode 3, the actual energy consumed is given by,
\begin{equation}
\bar{E}_i{'}= \sum_{\max(l,j)>R_i^*}\pi^1_{ml}\pi^2_{kj}P_i^*+ \sum_{\max(l,j)<=R_i^*} \left( \pi^1_{ml}\pi^i_{kj} \cdot \frac{2^{\max(l,j)}-1}{g_i} \right) \quad i=3
\end{equation}

Combining them together with optimal time fraction assigned for each mode, actual energy consumed per slot for this entire network, $\bar{E}_{act}$, is given by
\begin{equation}
\bar{E}_{act}=\sum_{i=1}^5 f_i^*\bar{E}_i^{'}. \label{eq:energy efficiency}
\end{equation}
We shall compare this actual energy consumption with the simulation result in the following section.

\section{Simulation}
We now present simulations to verify our findings and explore tradeoffs.  To ensure reasonable running times and memory requirements, we design for average arrival rates that are slightly larger than the actual rates, i.e., $\lambda_i \rightarrow (1+\epsilon)\lambda_i$ ($i=1,2$), where $\epsilon$ is a small positive value.
Noise at each node is assumed to be Gaussian with zero mean and unit variance and Rayleigh fading models are used for each link.

\begin{figure}[t]
   \centering
   \includegraphics[width = 12cm]{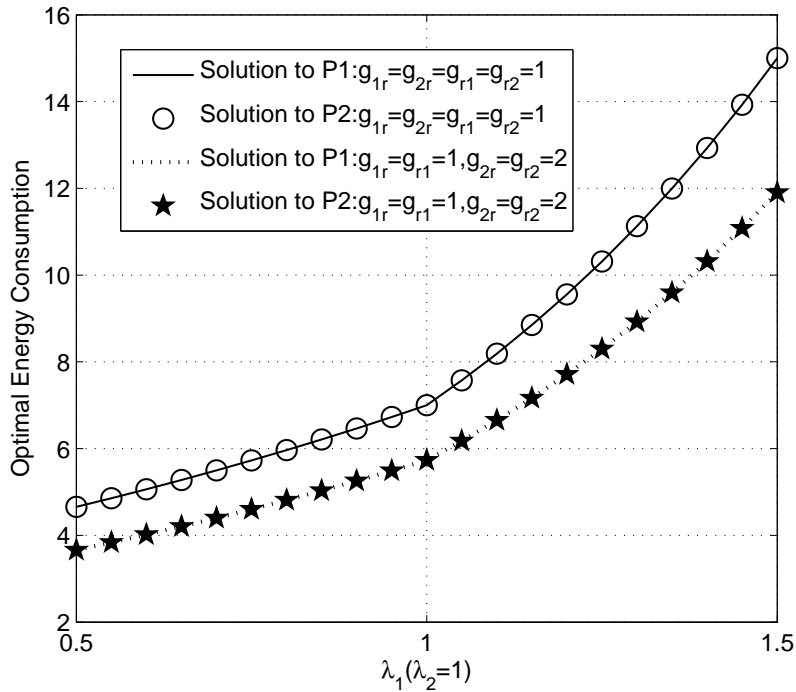}
   \caption{Optimal total energy consumption for some specified channel realizations for constant channel gains. $\lambda_2$ is set to be constant} \label{fig:energy_1}
   \end{figure}

In Fig. \ref{fig:energy_1}, the data arrival rate at $S_2$ is $\lambda_2=1$. An iterative sequential quadratic program (SQP) is used to solve {\bf P1} directly, and {\bf P2} is solved through a numerical solution of the KKT equations. It can be observed that the solution to \textbf{P1} and the solution to \textbf{P2} matches for different channel gain realizations and varying $\lambda_1$. The equivalence of problems \textbf{P1} and \textbf{P2} is therefore verified. In a similar manner, we have also numerically verified that P3 and P4 are equivalent, but omit the results in the interest of conciseness.

\begin{figure}[t]
   \centering
   \includegraphics[width = 12cm]{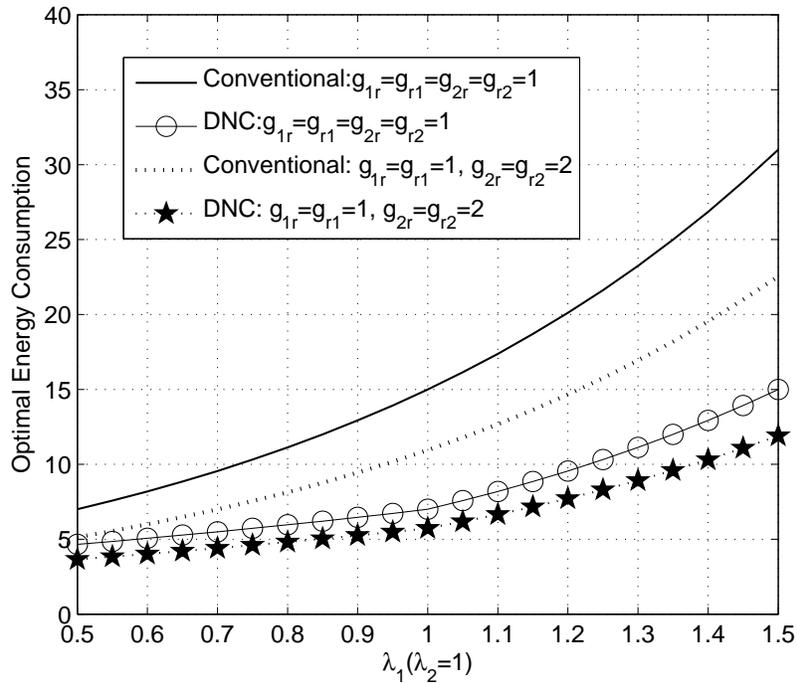}
   \caption{Optimal total energy consumption comparison of conventional scheme and digital network coding scheme for constant channel gains. $\lambda_2$ is set to be constant.} \label{fig:energy_2}
   \end{figure}

In Fig. \ref{fig:energy_2}, the energy efficiency of short term digital network coding and the conventional scheme (without network coding) are compared for some specified channel realizations. The energy required for supporting data arrival rate with conventional transmission is obtained by solving the problem below.
\begin{eqnarray}
    \min_{f_i} && \left( f_1P_1(\frac{\lambda_1}{f_1})+f_2P_2(\frac{\lambda_2}{f_2})+f_4P_4(\frac{\lambda_2}{f_4})+f_5P_5(\frac{\lambda_1}{f_5}) \right)\label{opt2}\\
     s.t. && f_i \geq 0 \\
         && \sum_{i=1}^5 f_i = 1  \label{Eqn:unity}\\
         && f_3 = 0
    \end{eqnarray}
It can be observed that energy efficiency is greatly improved by employing network coding. For instance, in the case of unit channel gains, energy is reduced from 31 units to 15 units, at $\lambda_1 = 1.5$.

\begin{figure}[!t]
   \centering
   \includegraphics[width = 12cm]{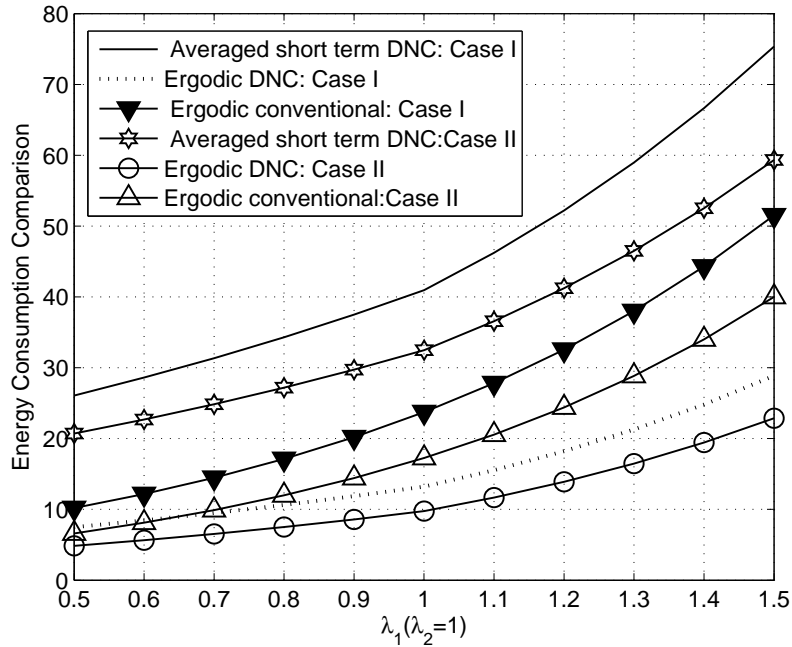}
   \caption{Optimal total energy consumption comparison for EERSP protocol. $\lambda_2$ is set to be constant. Case I denotes the case that $\overline{g}_{1r}=\overline{g}_{2r}=\overline{g}_{r1}=\overline{g}_{r2}=1$. Case II denotes the case that $\overline{g}_{1r}=\overline{g}_{r2}=1$ and $\overline{g}_{2r}=\overline{g}_{r1}=2$.} \label{fig:energy_4}
   \end{figure}

In Fig. \ref{fig:energy_4}, the energy consumption of the ergodic digital network
coding solution {\bf P3}, static-channel digital network
coding solution {\bf P1} and the ergodic conventional non-NC solution are compared, where the conventional non-NC solution is obtained by solving {\bf P3} with $f_3=0$. The
energy consumption of the static-channel DNC solution was obtained by averaging over 200 channel realizations.

It can be observed that energy efficiency is greatly improved by employing network coding. For instance, in the case of unit average channel gain, energy usage is reduced from over 50 units to less than 30 units, at $\lambda_1 = 1.5$. It is also observed that with network coding, optimizing over long term varying link gain can save even more energy, since it mitigates the adverse effect of deep fading scenario compared with short term DNC solution. Under the scenario of unit average channel gains, total energy consumed is reduced from over 70 units to less than 30 units, at $\lambda_1 = 1.5$.

\begin{figure}[!t]
   \centering
   \includegraphics[width = 12cm]{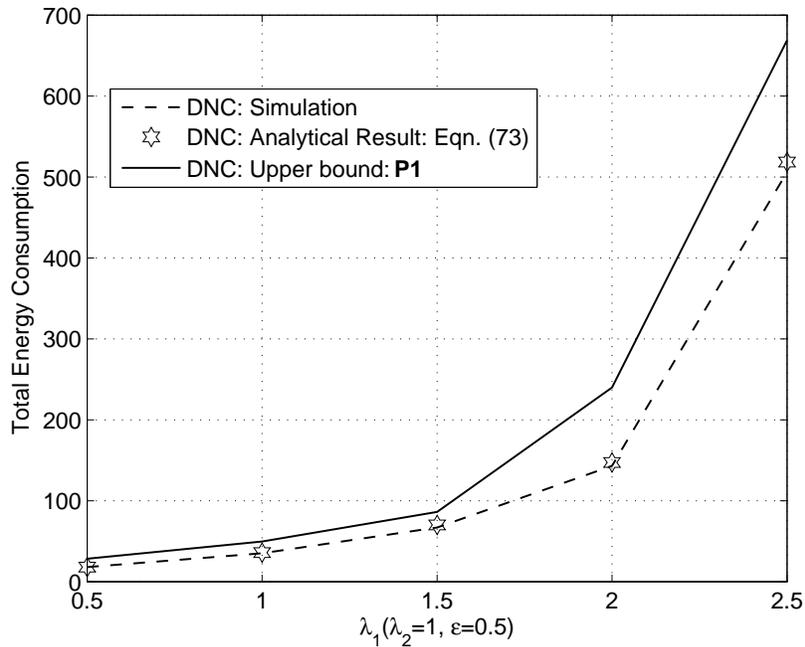}
   \caption{Actual total energy consumption comparison for EERSP protocol with positive $\epsilon$. $\lambda_2$ is set to be constant. The scenario of $g_{1r}
   =g_{r1}=1$,$g_{2r}=g_{r2}=2$ and $\epsilon=0.5$ is investigated.} \label{fig:energy_5}
   \end{figure}

We study the actual energy consumption by implementing EERSP protocol for the static-channel DNC in Fig. \ref{fig:energy_5}. In simulation, we ran $10^6$ slots and channels of each link are kept static throughout the simulation. It can be seen that the analytical result of actual energy usage matches well with the simulation result. As expected, the actual energy usage is less than the upper bound determined by the optimal energy minimization solution because the backoff parameter $\epsilon$ was non-zero.

%

   %

\begin{figure}[!t]
   \centering
   \includegraphics[width = 12cm]{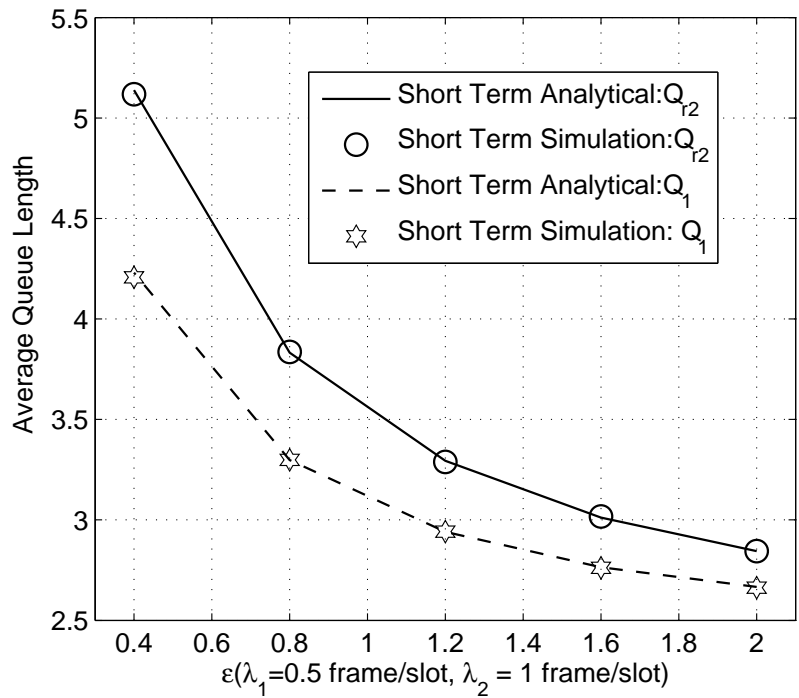}
   \caption{Average Queue Length at $S_1$  and $R$ for $S_2$ versus varying $\epsilon$ for {\bf P0}. All link gains are assumed to be unity.} \label{fig:queue_r}
   \end{figure}
   \begin{figure}[!t]
   \centering
   \includegraphics[width = 12cm]{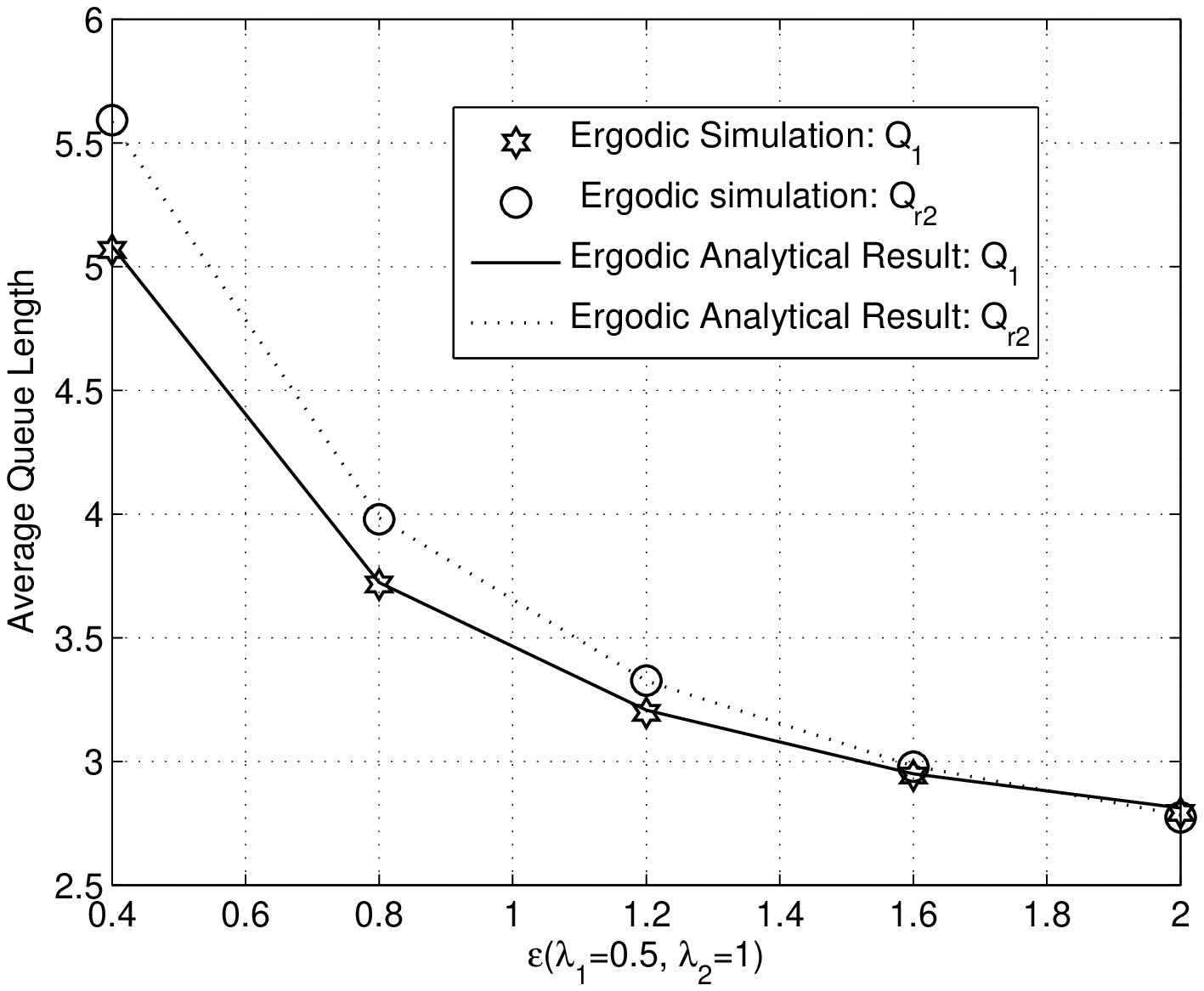}
   \caption{Average Queue Length at $S_1$ and $R$ for $S_2$ versus varying $\epsilon$ for {\bf P1}. All link gains are assumed to be unity on average.} \label{fig:queue_avg}
   \end{figure}

%

In the following, we show the tradeoff between energy efficiency and queue length. The data arrival process at each source node is assumed to be Poisson.
We also set $\lambda_1=0.5$ frame/slot and $\lambda_2=1$ frame/slot and each slot spans $1$ms.
The bandwidth is set to be 1kHz. All link gains are assumed to be unity and Gaussian noise at each node is assumed to
have zero mean and unit variance.
%

In Fig.\ \ref{fig:queue_r} and Fig.\ \ref{fig:queue_avg}, average queue length of the EERSP protocol for {\bf P1} and for {\bf P3} are shown respectively. It can be seen that the analytical result and simulation for queue length at both $S_1$ and at $R$ for $S_2$ match with each other perfectly. It is also observed that with increasing $\epsilon$, the queue length at $S_1$ decreases quickly due to higher transmission rate for each mode. However, increasing $\epsilon$ results in an increased probability of idle slots as we are over-provisioning the system, and therefore lower bandwidth efficiency. The factor $\epsilon$ thus controls the trade-off between bandwidth efficiency and buffering delay.

\section{Conclusion}
In this work, the minimal-energy allocation of resources to the five transmission modes in a two-way relay
network with digital network coding at the relay and stochastic packet arrivals at the two source nodes was
obtained. Both static and fading channels were accounted for, in the static-channel and ergodic energy
minimization problems respectively. The proposed solutions ensure stability of all four queues in the network
for arbitrary average packet arrival rates. The ergodic energy minimization solution has a water-filling
structure and lower energy usage compared to re-designing the system to match the instantaneous channel gains
using the static-channel energy minimization solution in a fading channel. In addition, a practical scheduling
protocol was introduced to implement the proposed resource allocation solutions, and an exact queuing analysis
of the protocol was obtained. As two-way relay networks appear in many applications including cellular
networks and satellite systems, the work presented here is an important step towards the realization of a
practical and useful communication network setup.

\appendices
\section{Proof for Proposition \ref{Lemma:7}}
In this appendix, we give the proof for Proposition \ref{Lemma:7}. Firstly, let us recall that the probability density function (pdf) of link gain over a Rayleigh fading channel is given by
\begin{equation}
p(g_i)=\frac{1}{\bar{g}_i}\exp(-\frac{g_i}{\bar{g}_i}). \label{eqn:dis}
\end{equation}

Hence, substituting (\ref{eqn:dis}) into (\ref{eqn:optR}), we can derive $\bar{R_i^*}$ as follows.
\begin{eqnarray}
\bar{R_i^*}&=&\int_{\frac{1}{\beta^*_i \log_2 e}}^\infty \log_2(\, \beta^*_ig_i \log_2 e  \,)\frac{1}{\bar{g}_i}\exp(-\frac{g_i}{\bar{g}_i})dg_i \\
&=& -\log_2(\, c_ig_i \,)\exp(-\frac{g_i}{\bar{g}_i})|_{c_i^*}^{\infty}+\int_{\frac{1}{c^*_i}}^\infty \frac{1}{g_i\ln 2}\exp(-\frac{g}{\bar{g}_i})dg_i\\
&=&\int_{\frac{1}{c^*_i}}^\infty \frac{1}{g_i\ln 2}\exp(-\frac{g}{\bar{g}_i})dg_i\\
&=&\int_{1}^\infty \frac{1}{g_i\ln 2}\exp(-\frac{g_i}{c_i^*\bar{g}_i})dg_i\label{eqn:app1}
\end{eqnarray}
Where $c_i^*=\beta^*_i \log_2 e$ for short. In the same manner, $\bar{P_i^*}$ is given by,
\begin{eqnarray}
\bar{P_i^*}&=&\int_{\frac{1}{c^*_i}}^\infty \frac{1}{g_i^2}\exp(-\frac{g_i}{\bar{g}_i})dg_i\\
&=&\int_{1}^\infty \frac{c_i^*}{g_i^2}\exp(-\frac{g_i}{c_i^*\bar{g}_i})dg_i\label{eqn:app2}.
\end{eqnarray}

Substituting (\ref{eqn:app1}) and (\ref{eqn:app2}) into (\ref{eqn:kkt1}), we obtain the following lemma.
\begin{lemma} \label{ergodicequality}
Under the optimal energy solution {\bf P4} over Rayleigh fading channels, optimal power level and optimal rate of each active mode can be linked by
\begin{equation}
\int_{1}^\infty (\frac{c_i^*}{g_i^2}-\frac{c_i^*}{g_i})\exp(-\frac{g_i}{c_i^*\bar{g}_i})dg_i+\gamma^*=0. \label{eqn:app3}
\end{equation}
\end{lemma}

Note that $c_i^*$ is an implicit function of $\bar{g_i}$. Take derivatives with respect to $\bar{g_i}$ on both sides of (\ref{eqn:app3}), we obtain
\begin{equation}
\frac{dc_i^*}{d\bar{g_i}}\frac{\int_{1}^\infty (\frac{1}{g_i^2}-\frac{1}{g_i})\exp(-\frac{g_i}{c_i^*\bar{g}_i})dg_i}{c_i^*}+\frac{d(\bar{g_i}c_i^*)}{d\bar{g_i}}\frac{\int_{1}^\infty (\frac{1}{g_i}-1)\exp(-\frac{g_i}{c_i^*\bar{g}_i})dg_i}{c_i^* \bar{g}_i^2}=0. \label{eqn:appequal}
\end{equation}

Since $\frac{1}{g_i^2}-\frac{1}{g_i}<0$ and $\frac{1}{g_i}-1<0$ hold if $g_i>1$, it can be deduced that $\frac{dc_i^*}{d\bar{g_i}}$ and $\frac{d(\bar{g_i}c_i^*)}{d\bar{g_i}}$ are with different signs, i.e., $\frac{dc_i^*}{d\bar{g_i}}\frac{d(\bar{g_i}c_i^*)}{d\bar{g_i}}<0$.

 Hence there are only two feasible scenarios: 1) $\frac{dc_i^*}{d\bar{g_i}}>0$ and $\frac{d(\bar{g_i}c_i^*)}{d\bar{g_i}}<0$; 2) $\frac{dc_i^*}{d\bar{g_i}}<0$ and $\frac{d(\bar{g_i}c_i^*)}{d\bar{g_i}}>0$. However, it can be immediately deduced that, if $\frac{dc_i^*}{d\bar{g_i}}>0$, $\frac{d(\bar{g_i}c_i^*)}{d\bar{g_i}}=c_i^*+g_i\frac{dc_i^*}{d\bar{g_i}}>0$ given $c_i^*,g_i>0$. A contradiction occurs for scenario 1) and it cannot be feasible. Therefore the latter scenario is uniquely feasible, i.e., $\frac{dc_i^*}{d\bar{g_i}}<0$ and $\frac{d(\bar{g_i}c_i^*)}{d\bar{g_i}}>0$.

Based on the above analysis, we seek to find the relation of $\bar{P}_i^*$ and $\bar{R}_i^*$ with $\bar{g}_i$ by deriving the first-order derivative of $\bar{P}_i^*$ and $\bar{R}_i^*$ with respect to $\bar{g}_i$
for each active mode.

From (\ref{eqn:app1}), the derivative of $R_i^*$ with respect to $\bar{g}_i$ can be given by,
\begin{eqnarray}
\frac{d\bar{R}_i^*}{d\bar{g}_i}&=&\frac{d(\bar{g_i}c_i^*)}{d\bar{g_i}} \cdot \frac{\int_{1}^\infty \frac{1}{\ln 2}\exp(-\frac{g_i}{c_i^*\bar{g}_i})dg_i}{(c_i^* \bar{g}_i)^2}>0 \label{derivative1}
\end{eqnarray}
where (\ref{derivative1}) comes from the conclusion that $\frac{d(\bar{g_i}c_i^*)}{d\bar{g_i}}>0$.
Therefore, $\bar{R}_i^*$ is an increasing function of $\bar{g}_i$ for each active mode in the ergodic case as well.

In a similar way, the derivative of $P_i^*$ in (\ref{eqn:app2}) with respect to $\bar{g}_i$ is derived as follows.
\begin{eqnarray}
\frac{d\bar{P}_i^*}{d\bar{g}_i}&=&\frac{dc_i^*}{d\bar{g_i}}\frac{P_i^*}{c_i^*} + \frac{d(\bar{g_i}c_i^*)}{d\bar{g_i}} \cdot \frac{\int_{1}^\infty \frac{1}{g_i}\exp(-\frac{g_i}{c_i^*\bar{g}_i})dg_i}{(c_i^* \bar{g}_i)^2}\label{eqn:derive1}\\
&=&\frac{dc_i^*}{d\bar{g_i}}(\frac{\int_{1}^\infty \frac{1}{g_i^2}\exp(-\frac{g_i}{c_i^*\bar{g}_i})dg_i \int_{1}^\infty (1-\frac{1}{g_i})\exp(-\frac{g_i}{c_i^*\bar{g}_i})dg_i}{\int_{1}^\infty (1-\frac{1}{g_i})\exp(-\frac{g_i}{c_i^*\bar{g}_i})dg_i} \nonumber\\
&&- \frac{\int_{1}^\infty \frac{1}{g_i}\exp(-\frac{g_i}{c_i^*\bar{g}_i})dg_i \int_{1}^\infty (\frac{1}{g_i}-\frac{1}{g_i^2})\exp(-\frac{g_i}{c_i^*\bar{g}_i})dg_i}{\int_{1}^\infty (1-\frac{1}{g_i})\exp(-\frac{g_i}{c_i^*\bar{g}_i})dg_i}  )\label{eqn:derive2}\\
&=&\frac{dc_i^*}{d\bar{g_i}}\frac{\int_{1}^\infty \frac{1}{g_i^2}\exp(-\frac{g_i}{c_i^*\bar{g}_i})dg_i \int_{1}^\infty \exp(-\frac{g_i}{c_i^*\bar{g}_i})dg_i-(\int_{1}^\infty \frac{1}{g_i}\exp(-\frac{g_i}{c_i^*\bar{g}_i})dg_i)^2}{\int_{1}^\infty (1-\frac{1}{g_i})\exp(-\frac{g_i}{c_i^*\bar{g}_i})dg_i}\\
&<&0\label{eqn:derive3}
\end{eqnarray}
where (\ref{eqn:derive2}) can be obtained by substituting the equality of (\ref{eqn:appequal}) into (\ref{eqn:derive1}) and (\ref{eqn:derive3}) comes from Cauchy--Schwarz inequality and the fact that $\frac{dc_i^*}{d\bar{g_i}}<0$.

Hence $\bar{P}_i^*$ is an increasing function of $\bar{g}_i$ for each active mode in the ergodic case as well. Proposition \ref{Lemma:7} then is proved.

\end{document}